%% file: elsarticle-template-num.tex
\documentclass[final,5p,times,twocolumn]{elsarticle} 

\usepackage{amssymb}
\usepackage{amsmath}
\usepackage{siunitx}
\usepackage{subcaption} 
\usepackage{color}
\usepackage{textcomp}
\usepackage{stfloats}

\usepackage{lineno,hyperref}
\modulolinenumbers[2]

\journal{Nuclear Inst. and Methods in Physics Research, A}

\begin{document}
	
	\begin{frontmatter}
		
		\title{Characterization of the photomultiplier tubes for the scintillation detectors of GRANDProto35 experiment}
			
		\author[1]{Xu Wang}
		\author[1]{Xiangli Qian\corref{cor1}}
		\ead{qianxl@sdmu.edu.cn}
		\author[2]{Yanhong Yu}
		\author[2]{Cunfeng Feng}
		\author[3]{Olivier Martineau-Huynh}
		\author[4]{Yi Zhang}
		\author[4]{Quanbu Gou}
		\author[4]{Wei Liu}
		\author[5]{Youliang Feng}

		\cortext[cor1]{Corresponding author}

		\address[1]{School of Intelligent Engineering, Shandong Management University, Jinan 250357, China}
		\address[2]{Institute of Frontier and Interdisciplinary Science, Shandong University, Qingdao 266237, China}
		\address[3]{Sorbonne Universit\'e, Universit\'e Paris Diderot, Sorbonne Paris Cit\'e, CNRS/IN2P3, LPNHE, Paris, France}
		\address[4]{Institute of High Energy Physics, Chinese Academy of Sciences, Beijing 100049, China}
		\address[5]{Key Laboratory of Dark Matter and Space Astronomy, Purple Mountain Observatory, Chinese Academy of Sciences, Nanjing 210008, China}
		
		\begin{abstract}
			GRANDProto35 is the first stage of the GRAND project. It will be composed of an array of 35 radio antennas and 24 scintillation detectors, the radio and scintillating subarrays being triggered independently. The scintillation detector array allow to cross check, through an offline treatment, if the selected radio candidates are indeed air shower events and thus quantitatively determine the detection efficiency of the radio array. The Hamamatsu R7725 is candidate for the scintillation detector photomultiplier. The characteristics of the PMT will directly affects the time and energy resolution, dynamic detection range of a scintillation detector. In order to cover the large dynamic range, voltage divider circuit featured with dual-readout was designed for the PMT. In this paper, details about the system setup, measurement method and results will be described. Some characteristics of PMT were calibrated and researched: the absolution gain, single photoelectron (SPE) energy resolution, transit time spread (TTS), linear dynamic range and temperature dependence of PMT gain.
			
		\end{abstract}

		
	\end{frontmatter}
	
	
	\input{introduction}

	\input{divider}
	\input{setups}
	\input{Characterization}
	\input{Conclusions}

	\section*{Acknowledgements} 
	We would like to thank the GRAND collaboration members for their valuable suggestions and support. This work was supported by the Natural Science Foundation of China (NSFC) (Grant No. 11705103) and the France China Particle Physics Laboratory.	
	
	\bibliographystyle{elsarticle-num}
	\bibliography{mydatabase}
	
\end{document}

%% file: introduction.tex
\section{Introduction}
The Giant Radio Array for Neutrino Detection (GRAND) is a proposed large-scale observatory to be built in China. Its ultimate goal is to discover and study the sources of ultra-high-energy (E$\geq$$10^{18}$eV) cosmic ray (UHECRs)~\cite{alvarez2020giant,Oliver_icrc2019}. The strategy of GRAND is to detect the radio emission emitted by large particle showers induced in the atmosphere by cosmic particles, with the largest array of radio antennas ever built. A staged construction ensures that key techniques are progressively validated, while simultaneously achieving important science goals. GRANDProto35~\cite{Gou2017} is the first stage of GRAND under construction, will lay the groundwork for future stages. It is located in the Tian Shan mountain in the Xinjiang province of China. It aims at demonstrating that radio-detection of air showers can be performed with very good background rejection, high efficiency, and an almost 100\% duty cycle.

\begin{figure}[htb]
	\centering
	\includegraphics[width=0.47\textwidth]{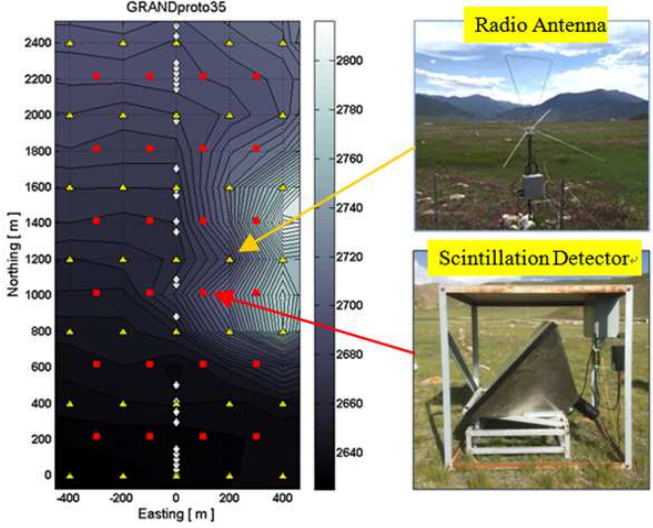}
	\caption{Left: the GRANDProto35 layout, which showing antennas (yellow triangles) and scintillation detectors (red squares), as well as the 21CMA pods (white diamonds). The color map shows the terrain elevation. Right: an antenna (top) and a prototype of scintillation detector (bottom).}\label{GP35_detectors}
	\vspace{0em}
\end{figure}

The GRANDProto35 detection units  are deployed on the infrastructure of the 21 CentiMeter Array (21 CMA)~\cite{Zheng2016Radio} and the Tianshan Radio Experiment for Neutrino Detection~\cite{charrier2019autonomous}, in a rectangular grid 800 m long in the East-West axis and 2400 m long in the North-South axis, as shown in Fig.~\ref{GP35_detectors}. The detection units are composed of 35 radio antennas and a particle detector array. The particle detector array is an autonomous surface array co-located with the antenna array. It consists of 24 scintillation detectors. If a radio EAS candidate is found to be in coincidence with a scintillator event composed of a signal in three detectors or more, then this guarantees with a $\sim$100\% probability that this event can indeed be associated with an EAS, as the probability for a random coincidence in such a large number of scintillators is negligible. Taking advantage of its high precision and efficiency, the scintillation detector will allow to determine the energy and direction of origin of the associated EAS. This will allow a precise determination of the radio EAS detection efficiency as a function of these two parameters. Finally a competitive scintillator array will turn GRANDProto35 into a multi-component detector that would also serve as platform for experimental technology tests and for the design of the future multi-messenger large scale observatory of GRAND.\\

The performance of the scintillation detector will directly affect the event reconstruction accuracy as said in the previous paragraph. As Fig.~\ref{GP35_detectors} show, a scintillation detector is equipped with a scintillation tile, light guide and a photomultiplier tube (PMT).  The scintillation tile is made of EJ-200 plastic scintillator~\cite{EJ-200} with a dimension of 70.7 cm $\times$ 70.7 cm $\times$ 3 cm. The scintillation light generated by the tile transfers to the photocathode of a PMT through the light guide. The characteristics of PMT affect the capacity of the detector greatly. A large dynamic range, a good linearity need to be achieved. Details of the requirements for the PMTs are summarized in Table \ref{table1}. The Hamamatsu R7725~\cite{R7725datasheet} was selected as the candidate for the scintillation detector, which is 2-inch diameter PMT with 12 dynode stages and standard bi-alkali photocathode. A dedicated voltage divider circuit for the dynode signal readout is designed to improve the linear dynamic range of the PMT. This paper describes the characteristics calibration of R7725: gain, linearity, time property and temperature dependence.
\begin{figure*}[hb]
	\begin{center}
		\includegraphics[width=0.8\textwidth]{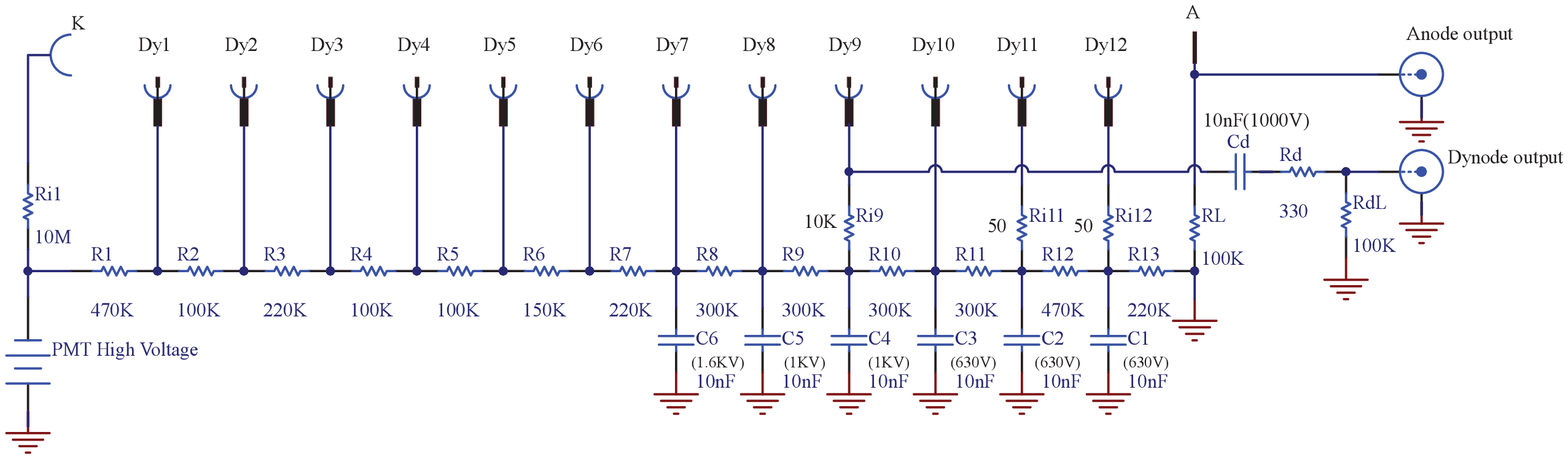}
		\caption{Schematic diagram of voltage divider circuit for R7725. All resistances are ohms and all capacitors are nanofarads.}
		\label{pmt_divider}
	\end{center}
\end{figure*}

%% file: divider.tex
\section{Voltage divider circuit of PMT}\label{voltage_divider}
The scintillation detector is designed to detect 1$\sim$1500 minimum ionized particles (MIP). The output signals of PMTs at working high voltage for
single MIP are approximately at 5 mV. Potential large signals will lead to the anode current of PMT going to saturation. In order to cover the high dynamic range with ideal linearity ability, voltage divider circuit featured with dual-readout was designed for PMTs: one from the anode, and the other from the $9^{th}$ dynode stage (Dy9).  The signal of anode is about 27 times that of Dy9 when PMT operates at working gain: $1.0\times10^{5}$. The schematic of the voltage divider circuit is presented in Fig.~\ref{pmt_divider}. It is a tapered divider, which employs higher resister values between the photocathode and Dy1 to improve signal-to-noise ratio. The decoupling capacitors at Dy7$\sim$Dy12 supply electric charge for the high pulses and restrain the voltage drop between the last dynode and the anode, resulting in an improvement of the linearity ability of PMT. Details about the anode and Dy9 linearity characteristics measurement results are presented in section \ref{section_linear}.

\begin{table}[h]\centering
	\caption{Requirements for main characteristics of PMTs to be used in the GRANDProto35 scintillation detector.}\label{table1}
	\vspace{1em}
	\begin{tabular*}{0.49\textwidth}{p{6cm}|c}\hline
		Parameter & Specified value\\\hline
		Photocathode diameter & $>$45 mm\\
		Quantum efficiency at 420 nm & $>$25\%\\
		Peak-to-Valley ratio & $>$1.5\\
		Transit Time Spread(FWHM) & $<$ 3 ns \\
		Dark count rate(0.3 spe threshold, at 25$^{\circ}$C) & $<$1000 Hz\\
		Late afterpulses between 100 ns and 10 $\mu$s & $<$15\%\\
		Anode pulse linearity at 5\% deviation & $>$50 mA \\\hline
	\end{tabular*}
\end{table}

%% file: setups.tex
\section{Experimental setups}\label{exper_setup}

\begin{figure}[htb]
\centering
\includegraphics[width=0.48\textwidth]{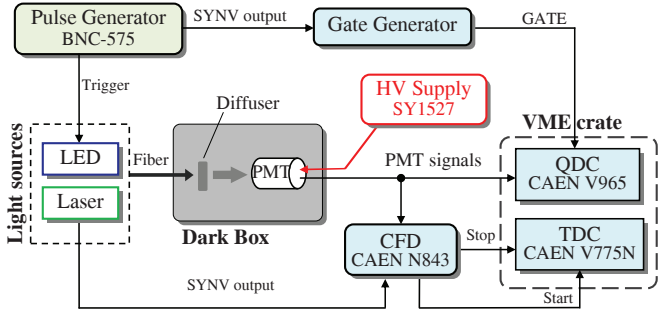}
\caption{Experimental setup used to measure PMT output charge and time properties.}\label{experimental_setup}
\vspace{0em}
\end{figure}

The experimental setup used for qualification and is shown in Fig.~\ref{experimental_setup}. It consists of light sources, a high voltage supply (CAEN SY1527), a pulse generator, dark box, trigger electronics and a series of VMEbus based digital electronics~\cite{Xu2016Setup}. The light flux of light sources enters the dark box through a fiber and is scattered by interposing a diffuser to make a uniform light illumination. There are two measuring modes: charge measurement and timing calibration. The system for charge measurement is depicted in the top half of the Fig.~\ref{experimental_setup}. The LED with a peak wavelength at $\sim$400 nm serves as the light source and is driven by a trigger signal from a pulse generator (BNC575) with pulse width of 30 ns. A synchronizing signal from the pulse generator is sent to a gate generator and the generated signal is supplied to the gate for QDC (CAEN V965).  The time width of the gate is set to 400 ns and the PMT signals are arranged within the gate by adjusting the relative time delay between trigger and synchronous signal. QDC V965 digitizes the output charge of PMT only within the time window, which depresses the random noises in a large level.

The bottom half of the Fig.~\ref{experimental_setup} shows the system setup for accurate time measurement. A picosecond pulse laser (Hamamatsu PLP-10) with a pulse width of 70 ps and a wavelength of 405 nm serves as the light source. Both PMT signals and the laser's synchronous output signal are fed into a 16-channel constant fraction discriminator (CFD: CAEN N843), the latter of which works as the reference logic start for the TDC COM signal. The CFD technique effectively increases the time accuracy by eliminating the time walk induced by amplitude dispersion through a fixed fraction of the full amplitude for signals' discrimination. The outputs of the CFD are then sent into a TDC (CAEN V775N) and it measures the relative time interval between the reference logic signal and PMT signals.

%% file: Characterization.tex
\section{Calibration and Characterization of the PMT}

All tests of PMTs are operated  at room temperature ($\sim$22 $^\circ$C) except for the measurement of temperature dependence of the gain. The head-on type PMTs are very sensitive to a magnetic field and in order to reduce the effects of the geomagnetic field, the PMTs are wrapped in magnetic shield cases. The testing setups used for all PMTs are described in section~\ref{exper_setup}. The measurement methods and results are shown in the following sections.

\subsection{Absolute gain and single photoelectron resolution}

An important parameter of a PMT is the absolute gain, which is required for the arrangement of the working high voltage and the calculation of the photoelectron number of MIP. In order to get the absolute gain of the PMT at a given voltage, the single photoelectron (SPE) spectrum is measured. To obtain an obvious SPE spectrum, the measuring high voltage is set to -1700V that is a little higher than the typical working voltage. The intensity of incident light is continuously weakened by turning down the driving amplitude for LED until most of the signals reach the level of the pedestal. On this single photoelectron light condition, the number of photoelectrons detected by the PMT can be described by the Poisson distribution:

\begin{equation}
P(n) = \frac{\lambda^n e^{-\lambda}}{n!}
\label{eq:1}
\end{equation}

where $\lambda$ is defined as the average number of photoelectrons collected by the first dynode and the P(n) is the  probability distribution of the number of n photoelectrons occurring in a given time period. Once most of the events are pedestals (P(0)), the single photoelectron events (P(1)) will form a high proportion of all signal events.  Based on this, the single photoelectron condition was reached via a quantized weight of the pedestal on all events.

\begin{figure}[htb]
\centering
\includegraphics[width=0.48\textwidth]{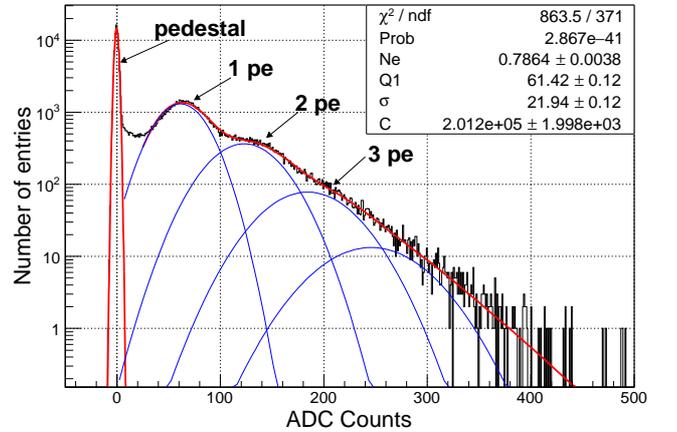}
\caption{The SPE charge spectrum in logarithmic view, together with the fit functions. The used ADC has a LSB of 0.03pC/count for the measurement.}\label{spe}
\vspace{0em}
\end{figure}

The SPE is a charge distribution spectrum that measured via QDC V965. The V965 is featured with dual gain stages and the low ($\times$1) gain stage was applied to increase the resolution when performing the measurement of SPE spectrum.  Fig.~\ref{spe}  shows a typical SPE spectrum test result of R7725. The first peak is the pedestal and is fitted with a Gaussian. The second peak is the SPE spectrum, which is fitted by convoluting a Poisson and a Gaussian fit~\cite{Bellamy1994_SPE}:

\begin{equation}
f(x)=C\sum_{n=0}^\infty \frac{(N_{pe})^ne^{-N_{pe}}}{n!\sqrt{2n\pi\sigma^{2}}}e^{\frac{-(x-nQ_1)^2}{2\pi\sigma^{2}}}
\label{eq:2}
\end{equation}

where C is the total number of the photoelectrons,  $N_{pe}$ stands for the number of expected average PE, $\sigma$ is the standard deviation of Gauss distribution, n denotes the indicate number of photoelectrons, $Q_1$ correspond to the charge of one photoelectron. The fit result shows that model and data are in good agreement. From the results of both fits the average charge Q corresponding to a SPE-signal is calculated
\begin{equation}
Q = \mu_{SPE}-\mu_{Ped}
\label{eq:3}
\end{equation}

$\mu_{SPE}$ and $\mu_{Ped}$ being the mean values of the fit function. The peak of the SPE is at 61.4 ADC counts and the calculated absolute gain is $1.3\times10^{7}$. The SPE spectrum also gives the energy resolution. Its standard deviation is a $\sigma$ of 21.9 and the relative number of $\sigma$/$Q_1$ is 0.35. The Peak/Valley ratio of SPE spectrum is around 2.9 for the R7725.

\subsection{Transit time spread (TTS)}
\begin{figure}[htb]
  \centering
  \includegraphics[width=0.45\textwidth]{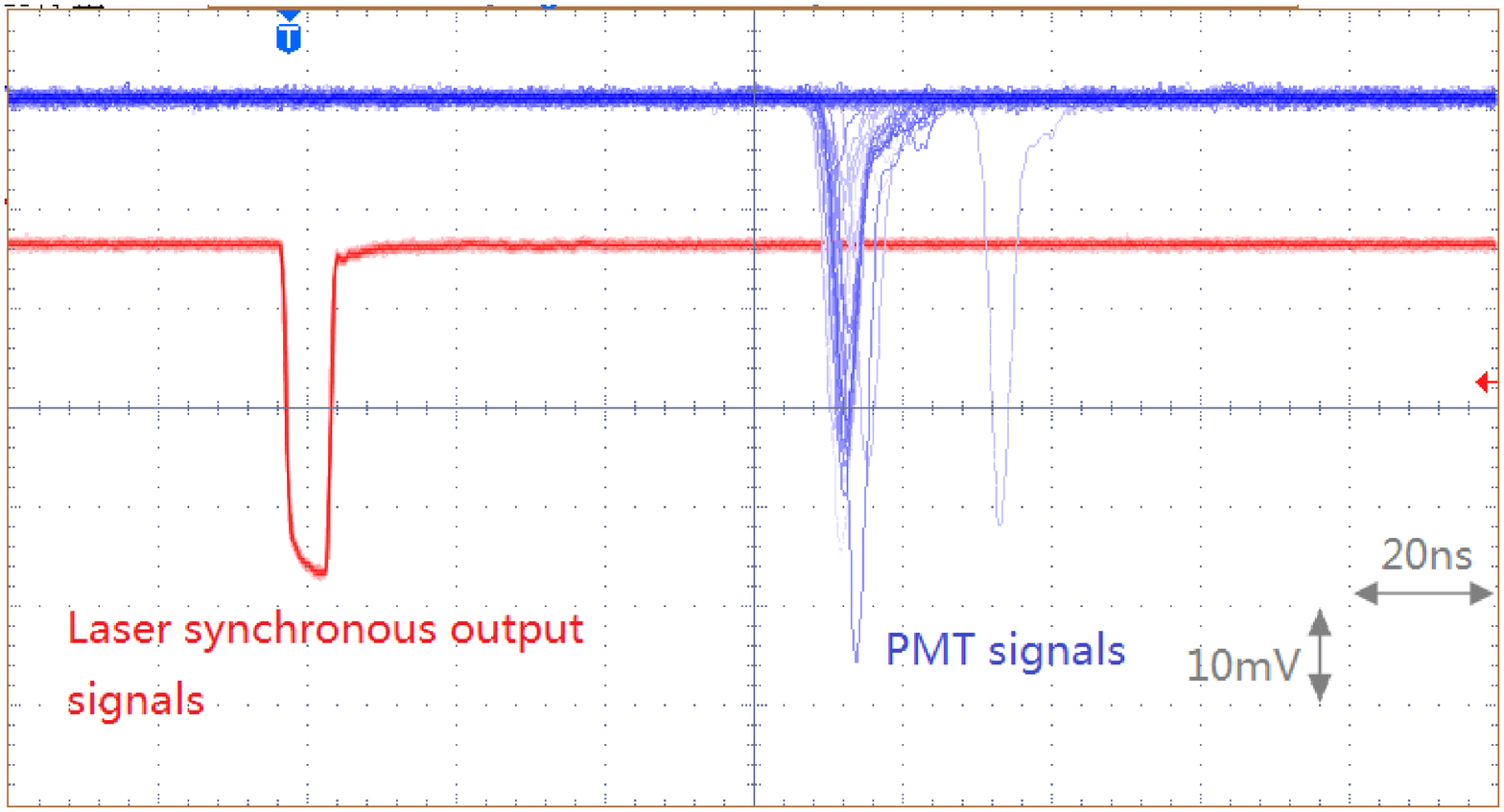}\\
  $(a)$\\
  \includegraphics[width=0.45\textwidth]{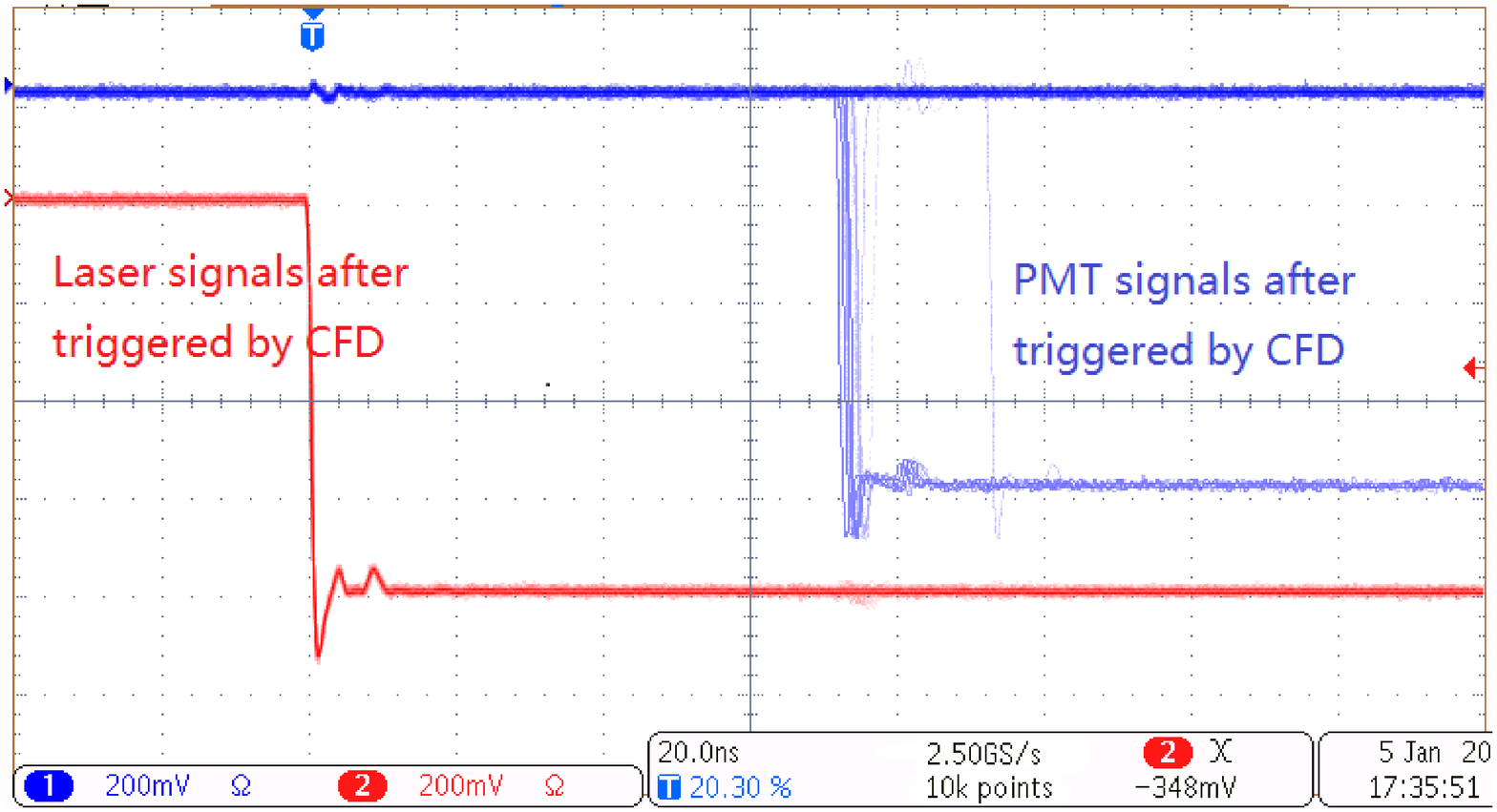}\\
  $(b)$\\
  \caption{(a): The PMT anode output pulses and laser synchronous output pulses. (b): The standard NIM signals after being triggered by the CFD. The waveform were captured by a digital Oscilloscope of 500 MHz bandwidth and 2.5 GS/s sampling rate.}\label{tts}
\end{figure}

The TTS determines the time resolution of the PMT and then affects the time resolution of the scintillation detector. The TTS has two main causes: the initial velocity spread of electrons emitted by different electrodes and the difference in transit time due to different points of emission from the same dynode~\cite{flycktphotomultiplier}. In the measurement, the TTS was measured as the time fluctuation of the intervals that between the synchronous signals of the laser and the PMT anode output signals. The system setup for the measurements of TTS was described in the second graph of section~\ref{exper_setup}. Fig.~\ref{tts} (a) shows the PMT anode output signals in SPE mode, as well as the laser's synchronous pulses and it was captured by a digital Oscilloscope with persistence display mode. It exhibits the transit time fluctuation of the PMT clearly. Fig.~\ref{tts} (b) shows the standard NIM output pulses that triggered by the CFD for the PMT and the laser's synchronous signals. Then the TDC measures the intervals between the leading edges of them.

\begin{figure}[htb]
  \centering
  \includegraphics[width=0.45\textwidth]{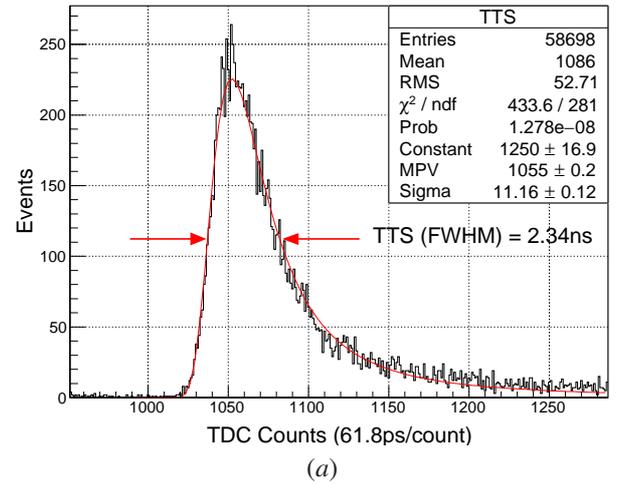}\\
  $(a)$\\
  \includegraphics[width=0.45\textwidth]{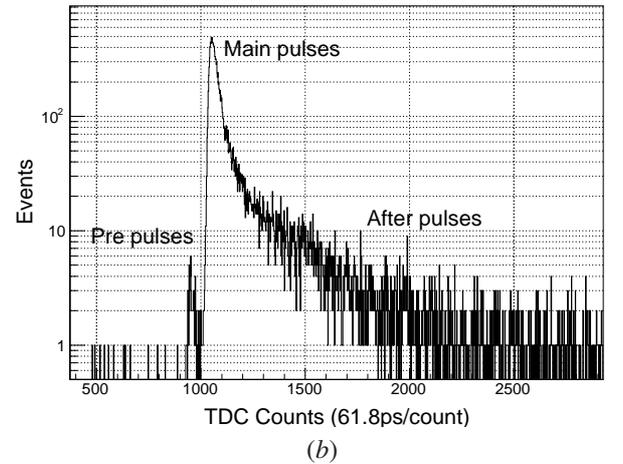}\\
  $(b)$\\
  \caption{(a): The distribution of TTS events together with fit function of Landau (red). (b): TTS spectrum in logarithmic view, witch displays the after pulse and pre-pulse clearly.}\label{tts at spe}
\end{figure}

The TTS is specified using the single photoelectron transit time spread. In the measurement, it begins with a pre-test of SPE spectrum to make sure it is at the SPE light illumination level. In this paper, the term TTS is understood as the full width half maximum (FWHM) of the distribution of transit time. Fig.~\ref{tts at spe} (a) shows the test results of TTS for R7725 with voltage supply of -1700V: the TTS value is 2.34 ns. It totally satisfies the physics requirement of the time resolution for the PMT. The TTS events present a Landau distribution and it exhibits a long tail, which shows obviously in Fig.~\ref{tts at spe} (b) with a logarithmic view. The tail is caused mainly by the low probability of after pulses. From our observation, the distribution gradually tends to a Gaussian as the increase of photoelectrons in the first dynode and the transit time fluctuation gets smaller for the PMT.

\subsection{Linearity of the anode and dynode signals}\label{section_linear}
\begin{figure}
  \centering
  \includegraphics[width=0.47\textwidth]{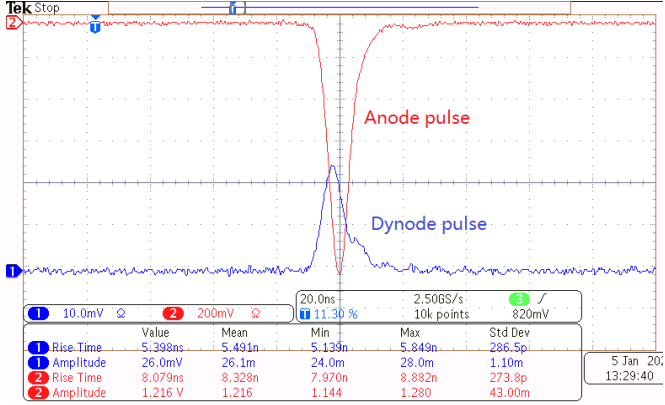}
  \caption{ PMT output pulse signals of anode (red) and Dy9 (blue), which was captured by an Oscilloscope. }\label{signal from osc}
\end{figure}

As described in section~\ref{voltage_divider}, to meet the requirement of wide dynamic linearity range for the experiment, a method of dynode readout was designed with a specific voltage divider circuit. The strategy of dual-readout (anode and Dy9) allows to avoid charge saturation with big pulse signals while increasing measurement resolution with small ones. Fig.~\ref{signal from osc} shows anode and dynode output signals of R7725 under -1200 voltage supply, which was captured with an oscilloscope. The dynode output is a positive pulse and the anode pulse is negative. Both the anode and Dy9 output signals should have good linearity behavior in their linear regions and a wide dynamic linear range needs to be generated by the implementation of Dy9 readout.

\begin{figure}
  \centering
  \includegraphics[width=0.44\textwidth]{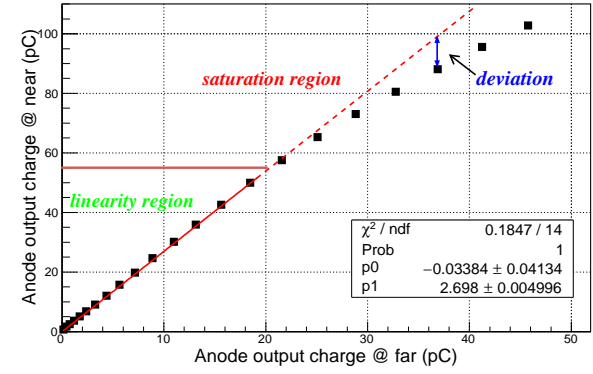}\\
  $(a)$\\
  \includegraphics[width=0.44\textwidth]{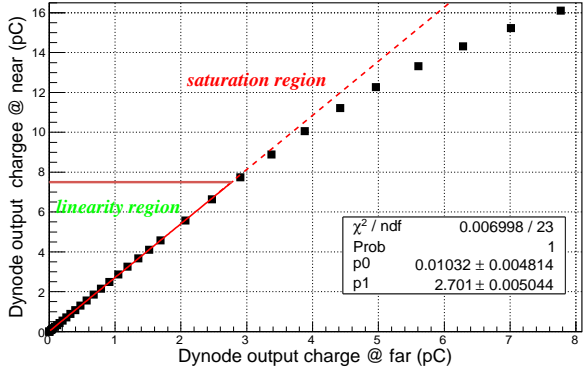}\\
  $(b)$\\
  \includegraphics[width=0.44\textwidth]{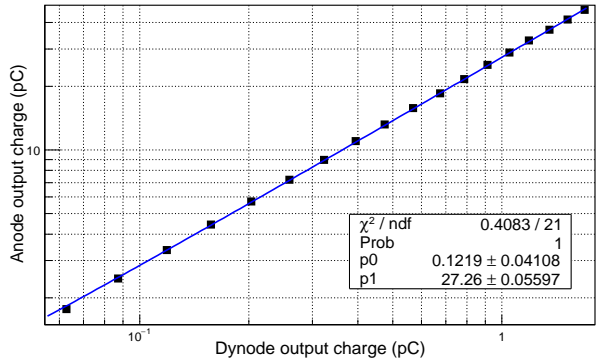}\\
  $(c)$\\
  \includegraphics[width=0.44\textwidth]{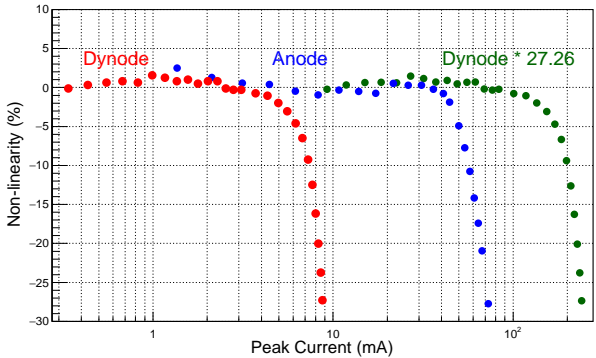}\\
  $(d)$\\
  \caption{Linearity of PMT R7725. (a) and (b) are the linear properties with strengthening of incident LED light illumination for the anode and Dy9 output signals. (c): The ratio of the anode output charge to the Dy9 output charge in the linear range. (d): The linear dynamic range for the anode and Dy9 output that calibrated with peak current.}
  \label{non-linearity}
\end{figure}

The linearity behavior of PMT is quantified by non-linearity, which is the degree of deviation from the linearity range. Non-linearity within $\pm$2\% is required for the PMTs in their linearity region and non-linearity larger than -5\% is considered as going into saturation. The non-linearity of anode and Dy9 were measured by the bi-distance method. The fiber of LED can be moved to a near or far position and the intensity of LED illumination will be gradually strengthened with increasing pulse voltage amplitude. Light intensity ratio $\lambda$ between near and far is constant. While a PMT works in a linear region, the ratio of PMT signals sizes is equal to $\lambda$ and it will get smaller than $\lambda$ when PMT going into saturation region.  Non-linearity is defined as:

\begin{equation}\label{eq linearity 1}
  \eta_i=({\frac{Q_i^{\prime}}{Q_i}-\lambda})/{\lambda}\times100\%
\end{equation}

where $\eta_i$ denotes the non-linearity that expressed as a percentage. $Q_i^\prime$ and $Q_i$ being the PMT's average output charges corresponding to near and far positions respectively. The parameter $\lambda$ can be given by a linear fitting of the measured spots $[Q_i, Q_i^\prime]$, where the fitting range is far from the saturation range. The linear fit function is shown as:

\begin{equation}\label{eq linearity 2}
  Q_i^{\prime}=p_0 \cdot Q_i + p_1
\end{equation}

where the linear fitted slope $p_0$ is $\lambda$. Fig.~\ref{non-linearity} shows the non-linearity measurement result of R7725 with gain $1.0\times10^5$ under HV supply -936.8V. The width of driving pulse for LED was set to 15 ns and under each measured point, the anode and Dy9 output charge and peak current were measured. Fig.~\ref{non-linearity} (a) and (b) are the ratios of PMT output charge under near and far position. In the linearity region, all testing points exhibit excellent linear properties. As the intensity of incident light is getting stronger, the measured spots gradually deviate from linearity, because of the space charge effect at the last few dynode stages, meaning that the PMT is going to saturation. Fig.~\ref{non-linearity} (c) shows the output charge ratio of anode and Dy9, which exhibits good linearity at a ratio value of 27.26 before saturation. Fig.~\ref{non-linearity} (d) shows the non-linearity percentage as a function of the observed peak current for anode and Dy9. The maximum currents correspond to a maximum charge is independent with the incident light pulse width.  Onset of saturation occurs, the measured peak anode current is used to calibrate the linearity dynamic range. The peak current is to absolutely calibrate the upper limit of the linearity dynamic range that is independent with the incident light pulse width. All test results about non-linearity are listed in Table~\ref{table}.

\begin{table}[h]\centering
\caption{The non-linearity test results.}\label{table}
\vspace{1em}
\begin{tabular*}{0.49\textwidth}{p{6.1cm}|c}\hline
Non-linearity characteristics & Value\\\hline
Measurement HV @ gain of $1.0\times10^5$ & -936.8 V \\
Anode \& Dy9 non-linearity in linear region & $<$ $\pm$2\% \\
Anode \& Dy9 peak  current & 50 mA / 6.3 mA\\
Dy9 peak current that converted into Anode & $\sim$ 170 mA \\
Anode / Dy9 output charge	& 27.26 \\\hline
\end{tabular*}
\end{table}

For the typical anode output, onset of saturation corresponding to a -5\% deviation from the linear region occurs for a 50 mA peak current under the working HV. The method of dual-readout that yield the Dy9 output increases it to around 170 mA. Consequently, the linear dynamic range of the scintillation detector can expand up to around 1700 MIP, which fully satisfies our physics requirements.

\subsection{Temperature dependence of PMT gain}

The long term stability of the scintillation detector is conditional upon the ambient temperature dependence of the PMT. In our measurement, a temperature coefficient is used to specify the temperature dependence. It is a combined coefficient that has two components: the photocathode sensitivity and the dynode multiplier gain. To calibrate the temperature coefficient, a  wide range of temperature from -30$^{\circ}$C to 40$^{\circ}$C is supplied by a temperature controlled chamber, which is approximately the same as the weather and climate changes in the station. The PMT was put into the chamber and was illuminated by the LED through 2 meters fiber. The duration of refrigeration is about 12 hours and the LED can't hold an extremely high stability in such a long time run. To eliminate testing error caused by the instability of LED, two PMTs that shared the same LED were operated synchronously in a stable room temperature and will supply a correction for the intensity of LED.  The voltage was adjusted to keep the PMT at the working gain of $1.0\times10^5$.

\begin{figure}[ht]
  \centering
  \includegraphics[width=0.48\textwidth]{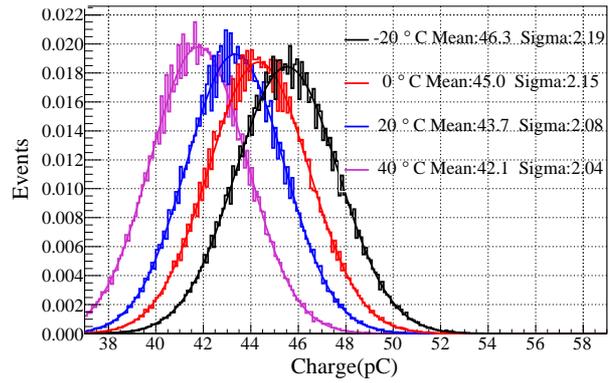}\\
  $(a)$\\
  \includegraphics[width=0.48\textwidth]{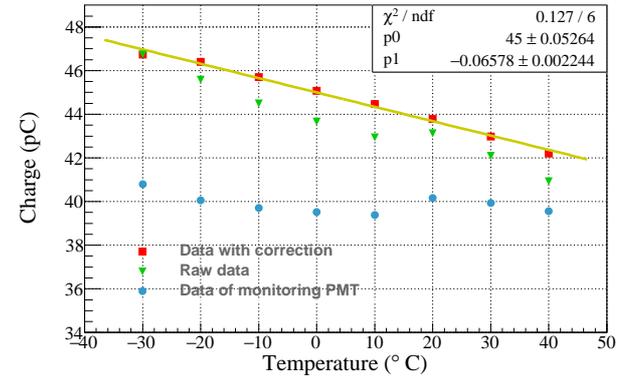}\\
  $(b)$\\
  \caption{(a): The charge distributions at temperatures: -20, 0, 20 and 40$^{\circ}$C, together with the Gaussian fitting line. (b): The PMT output charge versus temperature, as well as the linear fitting line (yellow). The triangle points are the measured raw data and the blue dot points are the data of monitoring PMT. The triangle red points are the results after the correction.}\label{temperature}
\end{figure}

Fig.~\ref{temperature} (a) shows the charge distributions at the temperatures: -20, 0, 20 and 40$^{\circ}$C. There is an obvious correlation between the gain and the operated temperature. The peaks have a decrease of approximately 10\% with a temperature rise of 60$^{\circ}$C. In order to get an accuracy temperature coefficient, the charge distributions under each operated temperature are fitted by Gaussian and all obtained mean values are drawn in Fig.~\ref{temperature} (b), which represent the output charge sizes of the PMT. It decreases from 46.7 pC at -30$^{\circ}$C to 42.1 pC at 40$^{\circ}$C and a linear temperature dependence of the charge size with negative slope is obtained.  The obtained coefficient is -1.46\textperthousand/$^{\circ}$C. It will be applied to correct the variation of the PMT gain caused by the ambient temperature, consequently improve the energy resolution of scintillation detector.

%% file: Conclusions.tex
\section{Conclusions}
The scintillation detector array will serve as the autonomous surface trigger array for GRANDProto35 and Hamamatsu R7725 was chosen as the candidate for the scintillation detectors. In order to cover the dynamic range of the physics detection for the detector, a divider circuit with function of dual-readout was designed for PMT. We have built test systems and operated effective method to study the properties of the PMT. Characteristics of the gain, linearity, TTS and temperature dependence of the gain were considered. The Hamamatsu R7725 showed good capacity of single PE energy and time resolution, which are in good agreement with the producer specifications. From the linearity measurement results, the additional dynode readout expands the dynamic linear range by a factor 3.4 and both the anode and dynode readout had ideal linear property in their linearity region. The measured temperature coefficient of PMT will serve as important reference for the correction of detected PE when perform the energy reconstruction.